\documentclass[aps,prb,superscriptaddress,reprint]{revtex4-1}
\usepackage{graphicx}
\usepackage{mathrsfs}
\usepackage{bm}
\usepackage{amsmath}
\usepackage{dcolumn}
\usepackage{epstopdf}
\usepackage{dsfont}
\usepackage{amssymb}
\usepackage{tabularx}
\usepackage{array}
\usepackage{color}
\usepackage[colorlinks=true, letterpaper=true, pdfstartview=FitV, linkcolor=blue, citecolor=blue, urlcolor=blue]{hyperref}

\providecommand{\U}[1]{\protect\rule{.1in}{.1in}}

\begin{document}

\title{Electrically modulated SQUID with single Josephson junction
coupled by a time-reversal breaking Weyl semimetal thin film}

\author{Yong Xu}
\affiliation{Department of Physics, South University of Science and Technology of China,
Shenzhen 518055, China}
\author{Salah Uddin}
\affiliation{Department of Physics, South University of Science and Technology of China,
Shenzhen 518055, China}
\author{Jun Wang}
\affiliation{Department of Physics, Southeast University, Nanjing
210096, China}
\author{Zhongshui Ma}
\affiliation{School of Physics, Peking University, Beijing 100871, China}
\affiliation{Collaborative Innovation Center of Quantum Matter, Beijing,
100871, China}
\author{Jun-Feng Liu}
\email{liujf@sustc.edu.cn}
\affiliation{Department of Physics, South University of Science and Technology of China,
Shenzhen 518055, China}

\begin{abstract}
Usually, the superconducting quantum interference device (SQUID) consists of
two Josephson junctions and the interference therein is modulated by a
magnetic flux. In this work, we propose an electrically modulated SQUID
consisting of single Josephson junction coupled by a time-reversal breaking
Weyl semimetal thin film. For a low Fermi energy, the Josephson current is only mediated by Fermi arc surface states, and has an
arbitrary ground-state phase difference $\varphi_0$ which is directly proportional to the product of the transverse electric field and the cross section area of the junction. For a suitable Fermi
energy, the bulk states make comparable contributions to the Josephson
current with the current-phase relation of a $0$-junction. The interference
between the surface channel and the bulk channel results in an electrically
modulated SQUID with single Josephson junction, which provides an experimental proposal to identify magnetic Weyl semimetals and may have potential
applications in superconducting quantum computation.
\end{abstract}

\maketitle

\section{Introduction}
As the host to Weyl fermions in condensed matter, the Weyl semimetal (WSM) is a topological semimetal where three-dimensional linearly
dispersed Weyl cones appear in pairs in momentum space \cite%
{wanxg11,aab11,aab112,xusy15,lvbq15,liujy15}. Two paired Weyl nodes have
opposite chiralities and are connected by Fermi arc surface states \cite%
{wanxg11,xusy15,okugawa14}. The essential property of Weyl fermions is the
apparent violation of charge conservation known as the chiral anomaly, which leads
to the unusual transport properties of WSMs, such as negative
magnetoresistance \cite{ninomiya83}, chiral magnetic effect \cite{chen13}, anomalous hall effect \cite{fang11}, and
non-local transport \cite{qixl13}. In these transport signatures, either
bulk states or surface states dominate the transport. Nevertheless,
unlike the fully gapped topological insulator, the gapless WSM hosts both
bulk states and surface states to support the transport, especially in the
thin film geometry. The investigation in the quantum interference between the bulk channel and
the surface channel in WSMs is very desirable.

On the other hand, since the recent experimental realization of Josephson $%
\varphi _{0}$-junction based on a nanowire quantum dot \cite{dbs16}, the interest in $\varphi
_{0}$-junctions has been revived \cite%
{liu16,wang16,alidoust16,malshukov16,chudnovsky16,feigelman17,loss17,bergeret17,vikstrom17,jin17}. The
so-called $\varphi _{0}$-junction, namely, the anomalous Josephson effect \cite%
{feinberg,buzdin,martin09,tanaka09,liu102,mints11,goldobin12,linder13,linder14,yokoyama14,klam14,rahnavard14,dolcini15},
has an unconventional current-phase relation (CPR) $I(\varphi )=I_{c}\sin (\varphi
-\varphi _{0})$, with an arbitrary ground-state phase difference $\varphi _{0}$. The tunable $\varphi _{0}$-junction has important applications in
superconducting computer memory components \cite{ecg15}, superconducting
phase batteries and rectifiers \cite{aar12}, as well as flux- or phase-based
quantum bits \cite{cp10}. Topological edge or surface states have also been
proposed to be employed to realize the $\varphi _{0}$-junction in two-dimensional or three-dimensional topological insulators where the bulk states are gapped \cite%
{tanaka09,dolcini15,wang16,alidoust16,jin17}. The WSM phase requires broken time-reversal (TR) or inversion
symmetry. Although the inversion symmetry breaking WSM has been experimentally identified \cite{xusy15,lvbq15,liujy15}, the evidence for TR breaking WSM is still lacking.
From the view of symmetry \cite{liu10}, the TR breaking WSM \cite{bernevig16} is a
natural platform to realize the anomalous Josephson effect. Although the
Josephson junction based on a WSM has been studied lately \cite%
{kundu16,kim16,kevin17,kundu17}, the anomalous Josephson effect has not been found. And only the bulk states are considered in these
studies. We will show that the transport via Fermi arc surface states can lead to a remarkable $\varphi _{0}$ state when a transverse electric field is applied to break some symmetry.

The interplay between bulk states and surface states is more interesting in the WSM based Josephson junction. In this work, we investigate the quantum interference between the bulk
channel and the surface channel in a TR breaking WSM thin film sandwiched between
two \textit{s}-wave superconductors. In the surface channel, electrons and
holes appear on the opposite surfaces. This spacial separation gives a chance to endow two paired electrons with different energies by a
transverse electric field, which leads to a tunable $\varphi _{0}$-junction
state. The bulk channel is not sensitive to the electric field and is
always a normal $0$-junction. The interference between surface states and bulk states results
in an electrically modulated superconducting quantum interference device
(SQUID) with single Josephson junction. The normal SQUID is usually modulated by a
magnetic flux and consists of two Josephson junctions. This
electrically modulated SQUID with single Josephson junction is a simple experimental proposal to identify the magnetic WSM, as well as a promising
platform for extensive applications in the fields of superconducting
electronics and superconducting quantum computation.

The rest of this paper is organized as follows. In Sec. II we introduce the model of the Josephson junction based on the Bogoliubov-de Gennes
equation and the tight-binding method, present the formula to calculate the Josephson current and ABS levels. In Sec. III we discuss the anomalous Josephson effect tuned by a transverse electric field when Fermi arc surface states dominate the transport. In Sec. IV, we discuss the SQUID effect stemming from the interference between surface states and bulk states. Finally, a brief summary is given in Sec. V.

\begin{figure}[tbp]
\begin{center}
\includegraphics[bb=129 0 759 589, width=3.415in]{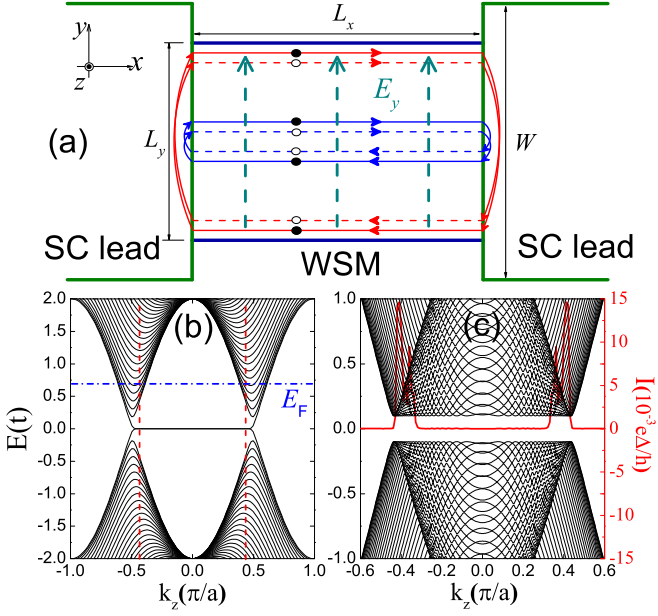}
\end{center}
\caption{(a) Schematic diagram of Josephson junctions linked by a TR
breaking WSM thin film between two \textit{s}-wave superconductors. The red
(blue) curves represent the Andreev bound states formed by Fermi arc surface
states (bulk states). (b) Energy dispersion $E(k_z)$ of the WSM thin film
with $L_y=50$ and $k_x=0$. Only states between two red dashed lines
contribute to the Josephson current for a fixed chemical potential ($\protect\mu_S=-4.4t$) in two
superconductors. (c) Quasiparticle excitation spectrum
of the \textit{s}-wave superconductor with parameters $W=60$, $\protect\mu%
_S=-4.4t$, $\Delta=0.1t$, and $k_x=0$. The red curve is the $k_z$ dependence
of the Josephson current with parameters $\Delta=0.01t$, $\protect\varphi=%
\protect\pi/2$ and $\protect\mu_W=0.1t$.}
\label{fig1}
\end{figure}

\section{Model and Formalism}
We consider a Josephson junction that
consists of a TR breaking WSM thin film sandwiched between two
general \textit{s}-wave superconductors. As shown in Fig. \ref{fig1}, the
hybrid junction lies along the $x$ direction and has a quantum constriction
in the $y$ direction. For simplicity, we assume that the translational
symmetry is preserved along the $z$ direction and thus the corresponding
wave vector $k_{z}$ is a good quantum number. In the normal state, the
TR breaking WSM is described by a minimal two-node model \cite%
{yang11}
\begin{eqnarray}
\mathcal{H_W} &=&\left( M-2t\sum\limits_{\alpha =x,y,z}\cos k_{\alpha }\right)
\sigma _{z}  \notag \\
&&+\lambda (\sin k_{x}\sigma _{x}+\sin k_{y}\sigma _{y})-\mu _{W}, \label{model}
\end{eqnarray}%
where $M=4t+2t\cos k_{0}$ determines the locations of two Weyl nodes $%
(0,0,\pm k_{0})$, $\sigma _{x,y,z}$ are the Pauli matrices for spin, $\lambda $
is the strength of the spin-orbit coupling, and $\mu _{W}$ is the chemical
potential in the WSM. The lattice constant is set to be $a=1$. To consider
a thin film geometry in the $y$ direction, we discretize the Hamiltonian
in real space along $x$ and $y$ directions. Then the discretized
Bogoliubov--de Gennes (BdG) Hamiltonian is
\begin{eqnarray}
H_{W} &=&\sum_{\mathbf{r},k_{z}}\Phi _{\mathbf{r},k_{z}}^{\dagger }\left(
\begin{array}{cc}
h_{w}(k_{z}) & 0 \\
0 & -h_{w}^{\ast }(-k_{z})%
\end{array}%
\right) \Phi _{\mathbf{r},k_{z}} \\
&&+\sum_{\mathbf{r},\mathbf{r}_{0},k_{z}}\left[ \Phi _{\mathbf{r}%
,k_{z}}^{\dagger }\left(
\begin{array}{cc}
h_{\mathbf{r}_{0}} & 0 \\
0 & -h_{\mathbf{r}_{0}}^{\ast }%
\end{array}%
\right) \Phi _{\mathbf{r+r}_{0},k_{z}}+H.c.\right] ,  \notag
\end{eqnarray}%
where $\mathbf{r}=(x,y)$ is the site index, $\mathbf{r}_{0}=\mathbf{x}$ or $%
\mathbf{y}$ represents the unit vector along $x$ or $y$ direction, $\Phi _{%
\mathbf{r},k_{z}}=[c_{\mathbf{r\uparrow },k_{z}},c_{\mathbf{r\downarrow }%
,k_{z}},c_{\mathbf{r\uparrow },-k_{z}}^{\dagger },c_{\mathbf{r\downarrow }%
,-k_{z}}^{\dagger }]^{T}$ is the field operator with $c_{\mathbf{r\uparrow (\downarrow )},\pm
k_{z}}$ the annihilation operator of an electron at site $\mathbf{r}$
with spin $\mathbf{\uparrow (\downarrow )}$ and momentum $\pm k_{z}$%
. The components included in the Hamiltonian are
\begin{eqnarray}
h_{w}(k_{z}) &=&(M-2t\cos k_{z})\sigma _{z}-\mu _{W},  \notag \\
h_{\mathbf{x}} &=&-t\sigma _{z}-\frac{1}{2}i\lambda \sigma _{x},h_{\mathbf{y}%
}=-t\sigma _{z}-\frac{1}{2}i\lambda \sigma _{y}.
\end{eqnarray}%
Moreover, a transverse electric field $E_{y}$ has also been considered
and modelled by linearly increasing on-site energies along the $y$ direction. It can be equivalently modelled by the modification of the chemical potential $\mu_W \rightarrow \mu_W-eE_yy$ with $e$ the unit charge.

For the two superconducting leads, we consider two general \textit{s}-wave
superconductors described by
\begin{eqnarray}
H_{S} &=&\sum_{\gamma ,\mathbf{r},k_{z}}\Phi _{\gamma ,\mathbf{r}%
,k_{z}}^{\dagger }\left(
\begin{array}{cc}
h_{s}(k_{z}) & \Delta e^{i\varphi _{\gamma }}i\sigma _{y} \\
\Delta e^{-i\varphi _{\gamma }}i\sigma _{y} & -h_{s}(k_{z})%
\end{array}%
\right) \Phi _{\gamma ,\mathbf{r},k_{z}} \\
&&+\sum_{\gamma ,\mathbf{r},\mathbf{r}_{0},k_{z}}\left[ \Phi _{\gamma ,%
\mathbf{r},k_{z}}^{\dagger }\left(
\begin{array}{cc}
t & 0 \\
0 & -t%
\end{array}%
\right) \Phi _{\gamma ,\mathbf{r+r}_{0},k_{z}}+H.c.\right] ,  \notag
\end{eqnarray}%
where $h_{s}(k_{z})=-2t\cos k_{z}-\mu _{S}$ with $\mu_S$ being the chemical potential in superconducting leads, the sum over $\gamma $ refers
to the left and right superconducting leads which are assumed to have the
same nearest-neighbor hopping energy $t$ as that in the WSM, $\Delta $ is
the superconducting gap, $\varphi _{\gamma }=\pm \frac{\varphi }{2}$ for the
left and right superconductor respectively with $\varphi $ the macroscopic
phase difference between two superconducting leads. The coupling between the
WSM and two superconducting leads is described by%
\begin{equation}
H_{C}=\sum_{\mathbf{r},k_{z}}\left[ \Phi _{\mathbf{r},k_{z}}^{\dagger
}\left(
\begin{array}{cc}
t & 0 \\
0 & t%
\end{array}%
\right) \Phi _{\mathbf{r+x},k_{z}}+H.c.\right] ,
\end{equation}%
where the sum over $\mathbf{r}$ refers to the left sites at the two
interfaces and two interfaces are assumed to be transparent for simplicity. Thus,
the whole Josephson junction is described by the Hamiltonian $%
H=H_{W}+H_{S}+H_{C}$. By using nonequilibrium Green's functions, the
Josephson current through column $l$ in the central WSM region for a
given $k_{z}$ is calculated by
\begin{equation}
I(k_{z})=\frac{1}{h}\int_{-\infty }^{\infty }\text{Tr}\left[ \check{t}^{\dag
}\check{e}G_{l,l-1}^{<}(k_{z})-\check{e}\check{t}G_{l-1,l}^{<}(k_{z})\right]
dE,  \label{SUPERI}
\end{equation}%
where $\check{t}=-t\tau _{3}\otimes {\sigma }_{z}+\frac{1}{2}i\lambda \tau
_{0}\otimes \sigma _{x}$ and $\check{e}=-e\tau _{3}\otimes {\sigma }_{0}$
denote the hopping matrix and the charge matrix respectively. $\tau _{3}$ ($%
\tau _{0}$) is the Pauli (unit) matrix in Nambu space. In equilibrium, the lesser-than Green's
function is calculated by $G^{<}=f\left( E\right) \left[ G^{a}-G^{r}\right] $%
,$\ $where $f\left( E\right) $ is the Fermi-Dirac distribution function. The
retarded and advanced Green's functions read
\begin{equation}
G^{r}(E)=[G^{a}(E)]^{\dag }=\frac{1}{E-H_{D}-\Sigma _{L}^{r}(E)-\Sigma
_{R}^{r}(E)},
\end{equation}%
where $H_{D}$ is the Hamiltonian of the WSM region. The retarded
self-energy $\Sigma _{L(R)}^{r}(E)$ due to coupling with the
superconducting leads L(R) can be calculated numerically by the recursive
method. Finally, the total Josephson current is given by $J=\frac{L_{z}}{%
2\pi }\int_{-\pi /a}^{\pi /a}I(k_{z})dk_{z}$.

In addition, the Andreev bound state (ABS) spectra can also be numerically
calculated through the Green's function technique. The ABSs result in peaks of particle
density within the superconducting gap. By searching the peaks of
particle density in column $l$ $\left( L_x\geqslant l\geqslant 1\right) $
\begin{equation}
\rho _{l}=-\frac{1}{\pi }Im\left[ \text{Tr}\left\{ G^{r}\left( l,l\right)
\right\} \right]  \label{SABS}
\end{equation}%
at a given phase difference $\varphi $, the energies of ABS levels can be located.
Then the ABS spectra can be obtained by scanning $\varphi $, which is helpful for
understanding the behavior of Josephson current.

\begin{figure}[tbp]
\begin{center}
\includegraphics[bb=10 254 673 1214, width=3.215in]{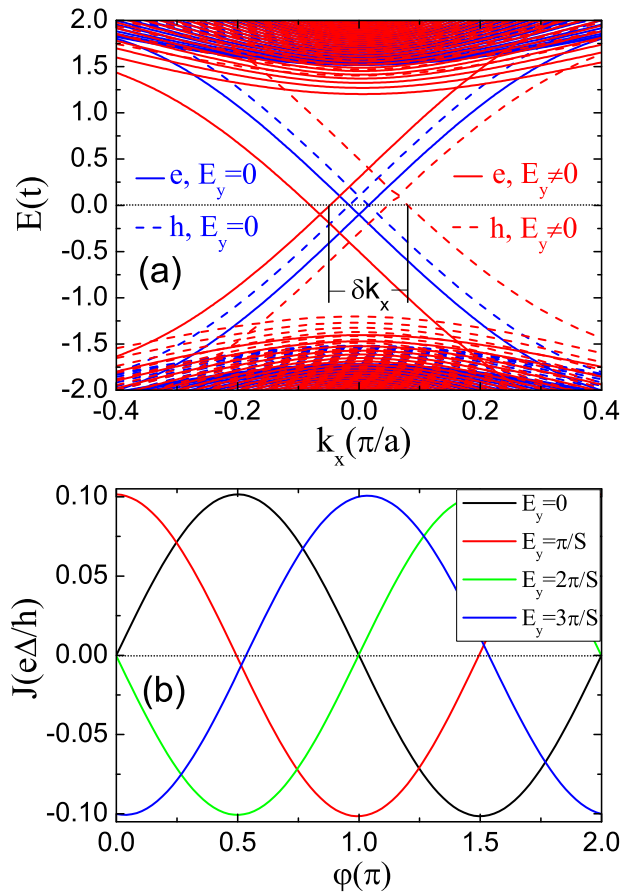}
\end{center}
\caption{(a) Energy dispersion $E(k_x)$ for electrons (solid lines) and
holes (dashed lines) in the WSM without (blue lines) or with (red lines) a
transverse electric field $E_y=0.008$, $k_z=0.2\protect\pi$. (b) Anomalous
Josephson effect with tunable ground-state phase differences for different values of $E_y$ which varies from $0$ to $3\protect\pi/S$ with cross section area $S=L_xL_y$. The
temperature $T=0.5T_c$, where $T_c$ is the critical temperature. The common
parameter is $\protect\mu_W=0.1t$.}
\label{fig2}
\end{figure}

\begin{figure}[tbp]
\begin{center}
\includegraphics[bb=66 48 556 573, width=3.415in]{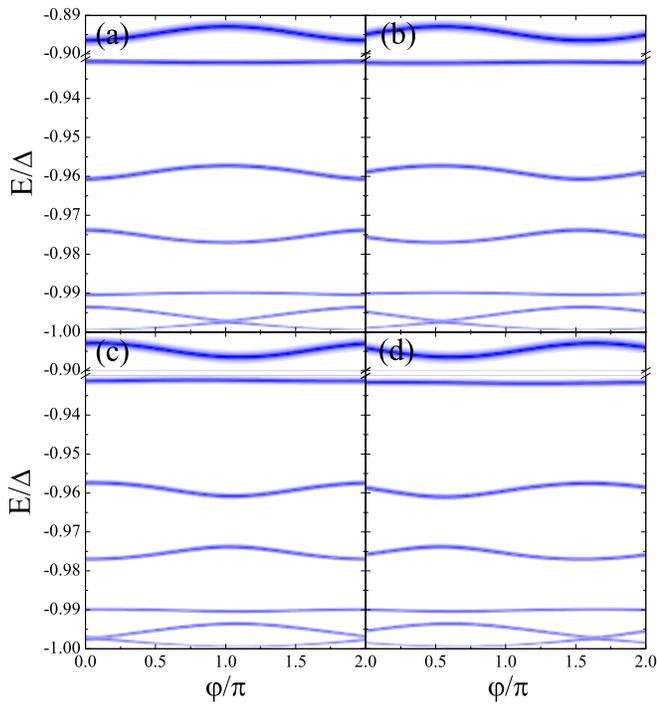}
\end{center}
\caption{ABS spectra with fixed $k_z=0.42\pi$ and various transverse electric fields (a) $E_y=0$, (b) $E_y=\pi/S$, (c) $E_y=2\pi/S$, and (d) $E_y=3\pi/S$.
 Other parameters are the same as those in Fig.
\protect\ref{fig2}.}
\label{abs}
\end{figure}

\section{Anomalous Josephson Effect}
Next, we present the numerical
results for the Josephson current. In our numerical calculations, $t=1$ is
the unit of energy, $\lambda =2$ and $\Delta =0.01$. $a=1$ is the unit of length, $%
1/a$ is the unit of the wave vector and $k_{0}=0.5\pi $. The geometric
parameters of the junction are set to be $L_{x}=100$, $L_{y}=50$, $%
L_{z}=1000 $, and $W=100$. The unit of transverse electric field $E_{y}$ is set to be $t/ea$ while a constant chemical potential ($\mu _{S}=-4.4t$) is used for the two superconductors. The range of $k_{z}$ of the electronic states in
the Fermi surfaces is determined by $\mu _{S}$, \textit{i.e.}, $\left\vert k_{z}\right\vert <0.43\pi$, is
consistent with the $k_{z}$ range in which the Josephson current is nonzero
(as shown in Fig. \ref{fig1} (c)).

First, we consider the situation where the chemical potential in the WSM is low, for
example, $\mu _{W}=0.1t$. For such a low $\mu _{W}$, there exist only Fermi
arc surface states in the range $\left\vert k_{z}\right\vert <0.43\pi $ (see
Fig. \ref{fig1} (b)). For each given $k_z$, the WSM is mapped to a two-dimensional quantum anomalous Hall (QAH) insulator. The QAH edge states  are responsible for the so-called Fermi arc surface states. The spin texture of the QAH edge states stemmed from this WSM model (Eq. (\ref{model})) is shown \cite{jiansheng14} to permit the Andreev reflections between the edge states at the upper and the bottom edges. It means that the Fermi arc surface states can form ABSs.
As sketched in Fig. \ref{fig1} (a), in such ABSs, electrons are localized in one surface while
holes in the other surface. The separation of electrons and
holes in space makes it possible that a transversal electric field $E_{y}$
endow two paired electrons with different energies. Thus electrons and
holes have different wave vectors.

Fig. \ref{fig2} (a) shows the $E_{y}$
induced wave vector difference $\delta
k_{x}=k_{x}^{e}-k_{x}^{h}=-E_{y}L_{y}/\lambda $. In the formation of ABS, this difference in wave vector leads to an additional
phase accumulation $\delta k_{x}L_{x}$ (for the right-going ABS) due to the travelling of
electrons and holes. This additional phase
should be offset by the phase difference of two superconductors $%
\varphi $. Therefore, the phase shift, or the ground-state phase difference
will be $\varphi _{0}=\delta k_{x}L_{x}=-E_{y}S/\lambda $ with $S=L_{x}L_{y}$%
, which is consistent with the CPRs shown in Fig. \ref{fig2} (b) for $%
\lambda =2$. The temperature is taken to be $T=0.5T_c$, which ensures that the first harmonic dominates the CPR. Moreover, numerical results of ABSs (shown in Fig. \ref{abs} verify the same $E_{y}$-induced phase shift $\varphi _{0}$. We can see that ABS spectra move left with increasing $E_y$.

It is noticeable that the transverse electric field is necessary to realize a $\varphi_0$-junction from the view of symmetry. When the electric field is absent, the WSM has a combined symmetry $R_y\sigma_x \mathcal{T}$ ($R_y$ is the reflection in $y$ direction and $\mathcal{T}$ is the time-reversal) which forbids the anomalous Josephson effect \cite{liu10}.

\begin{figure}[tbp]
\begin{center}
\includegraphics[bb=0 17 827 547, width=3.415in]{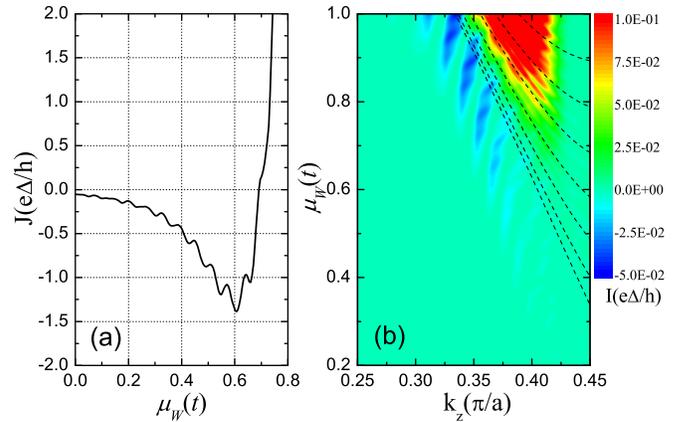}
\end{center}
\caption{The Josephson current as a function of $\protect\mu_W$ with fixed $%
E_y=2\protect\pi/S$. (a) $J(\protect\varphi=\protect\pi/2)$ versus $\protect\mu%
_W$. (b) Contour plot of $I(\protect\varphi=\protect\pi/2)$ versus $\protect\mu%
_W$ and $k_z$. The dashed curves are the energy dispersions of electrons in
the WSM as a reference. Other parameters are the same as those in Fig.
\protect\ref{fig2}.}
\label{fig3}
\end{figure}

\section{Electrically Modulated SQUID}
 When $\mu _{W}$ increases, the
bulk states gradually participate in the transport in the range $\left\vert
k_{z}\right\vert <0.43\pi $. As a result there are two channels available to carry the
Josephson current, one is the surface channel and the other is the bulk channel.
These two channels form an electrically modulated SQUID. For a suitably chosen $%
\mu _{W}$, the two channels can have comparable contributions to the
supercurrent. To find this suitable value of $\mu _{W}$, we set the surface channel
to be a $\pi $-junction by setting $E_{y}=2\pi /S$. Since the supercurrent
from the bulk channel is not sensitive to $E_{y}$ and remains always a $0$%
-junction, the supercurrents from two channels will cancel each other. At a
suitable $\mu _{W}$, the total Josephson current vanishes. Fig. \ref{fig3} (a)
shows the total Josephson current as a function of $\mu _{W}$ when the
phase difference is fixed to $\varphi =\pi /2$. The Josephson current first
decreases gradually with increasing $\mu _{W}$ because the supercurrent in the
surface channel is greatly enhanced due to the larger penetration depth of
the surface states. The penetration depth sensitively determines the
coupling of electron and hole, thus the amplitude of Andreev reflection and
Josephson current. For a higher $\mu _{W}$, the bulk channel also participate in the transport, the Josephson current goes up sharply, and approaches to $0$ nearly at $\mu
_{W}=0.69t$. In addition, the oscillations in the supercurrent come from the
multi-reflection in the normal reflection at interfaces. Fig. \ref{fig3} (b)
shows the $k_{z}$ resolved supercurrent $I(k_{z},\varphi =\pi /2)$ as a
function of $\mu _{W}$. It is clearly shown that the bulk channel is open at lower $\mu _{W}$ for larger $k_{z}$, which is consistent with
the energy dispersion of electrons in the WSM.

\begin{figure}[tbp]
\begin{center}
\includegraphics[bb=5 6 837 547, width=3.415in]{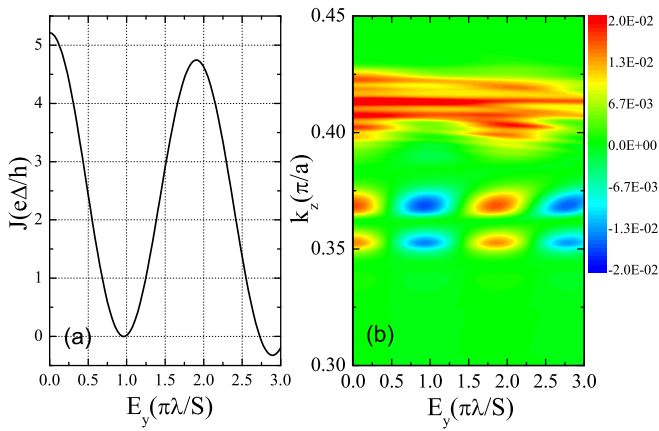}
\end{center}
\caption{The Josephson current as a function of $E_y$ with fixed $\protect\mu%
_W=0.69t$. (a) $J(\protect\varphi=\protect\pi/2)$ versus $E_y$. (b) Contour
plot of $I(\protect\varphi=\protect\pi/2)$ versus $E_y$ and $k_z$. Other
parameters are the same as those in Fig. \protect\ref{fig2}.}
\label{fig4}
\end{figure}

At $\mu _{W}=0.69t$, the two channels almost contribute the same amplitude
of the supercurrent. Since the surface channel is a $\varphi _{0}$-junction and the bulk channel remains a $0$-junction, the total Josephson current in the first harmonic approximation is expected to be
\begin{eqnarray}
J &=&J_{0}[\sin (\varphi -\varphi _{0})+\sin \varphi ]  \notag \\
&=&2J_{0}\cos \frac{\varphi _{0}}{2}\sin (\varphi -\frac{\varphi _{0}}{2}),
\label{jpr}
\end{eqnarray}%
where $\varphi _{0}=-E_{y}S/\lambda $. The critical current is defined to be $%
J_{c}=2J_{0}\left\vert \cos \frac{\varphi _{0}}{2}\right\vert
=2J_{0}\left\vert \cos \frac{E_{y}S}{2\lambda }\right\vert $. In particular,
the Josephson current at $\varphi =\frac{\pi }{2}$ is $J(\frac{\pi }{2}%
)=J_{0}[1+\cos \varphi _{0}]$, which gives a good fitting of our numerical
results shown in Fig. \ref{fig4} (a). The small deviation is due to the
slight decrease of bulk supercurrent with increasing $E_{y}$. Fig. \ref{fig4}
(b) clearly shows that the surface supercurrent is periodically modulated by
$E_{y}$ while the bulk supercurrent is not sensitive to $E_{y}$.

As shown in Eq. \ref{jpr}, the phase shift $%
\varphi _{0}$ in the surface supercurrent  is directly proportional to the transverse electric field $%
E_{y}$ and the cross section area $S=L_{x}L_{y}$, which is similar to the
situation in the usual magnetically modulated SQUID. Now we comment on the
conditions in which this simple relation is valid. First, the surface states
should be localized enough to the surfaces. Otherwise, the effect of $E_{y}$
will be weaker. It means that the $k_{z}$ range should keep away enough from
the Weyl nodes. The key parameter to make this condition satisfied is $\mu
_{S}$ which determines the $k_{z}$ range. The second condition is $\mu
_{W}\ll 2t(1-\cos k_{0})$ which makes the dispersion $E(k_{x})$ of surface
states linear and the Fermi velocity remains $\lambda .$

Finally, we comment on the experimental realization of the modulation of the transversal electric field. First, two gate voltages at two surfaces of WSM can induce an exactly transverse electric field. Second, even in the presence of longitudinal component of the electric field, the Josephson current will not change much based on the following considerations. The ABSs formed by Fermi-arc surface states separate electron and hole in space only along the y direction. Therefore only the $y$ component of electric field $E_y$ can endow two paired electrons with different energies, thus endow electron and hole with different wave vectors. It is just this wave vector difference between electron and hole that leads to an anomalous phase shift, and finally results in the oscillation of the critical current from the interference with the bulk ABSs. The numerical results also verify that the other components of electric field do not affect the Josephson current much.

\section{Conclusion}
In conclusion, we propose an electrically modulated
SQUID with single Josephson junction coupled by a TR breaking
Weyl semimetal thin film. There exist two channels, the surface channel and
the bulk channel, to carry the supercurrent. The surface channel serves as a
$\varphi _{0}$-junction where the ground-state phase difference is simply
modulated by a transverse electric field as $\varphi _{0}=-E_{y}S/\lambda$%
. The bulk channel remains always a $0$-junction. The quantum interference
between the two channels results in an electrically modulated SQUID. This proposed Josephson junction with arbitrarily tunable critical current and ground-state phase difference may have potential applications in the fields of superconducting electronics and
superconducting quantum computation.

\begin{acknowledgments}
The work described in this paper is supported by the National Natural
Science Foundation of China (NSFC, Grant Nos. 11774144, and
11274059).
\end{acknowledgments}

\end{document}